\documentclass[apj]{emulateapj}
\begin{document}
\title[MHD turbulence]{Spectra of strong MHD Turbulence from high-resolution simulations}
\author{Andrey Beresnyak}
\affil{Los Alamos National Laboratory, Los Alamos, NM, 87545}

\bibliographystyle{apj}

\begin{abstract}
 Magnetohydrodynamic turbulence is a ubiquitous phenomenon in solar physics, plasma physics
 and astrophysics and governs many properties of the flows of well-conductive fluids.
 Recently, conflicting spectral slopes for the inertial range of MHD turbulence has been reported
 by different groups. Varying spectral shapes from earlier simulations hinted at a wider spectral
 locality of MHD, which necessitated higher resolution simulations and careful and rigorous numerical
 analysis. In this Letter we present two groups of simulations with resolution up to $4096^3$
 that are numerically well-resolved and has been analyzed with exact and well-tested method
 of scaling study. Our results from both simulation groups indicate that the power spectral
 slope for all energy-related quantities, such as total energy and residual energy are around $-1.7$,
 close to Kolmogorov's $-5/3$. This suggests that residual energy is the constant fraction, $0.15\pm0.03$
 of the total energy and that in asymptotic regime magnetic and kinetic spectra have the same scaling.
 The $-1.5$ slope for energy and $-2$ slope for residual energy
 suggested by other groups seems to be completely inconsistent with numerics.     
\end{abstract}


\maketitle
\section{Introduction} 
Most astrophysical, stellar and space plasmas are well-ionized and well-conductive. On large scales
they are often described as an ideal MHD fluid -- a perfectly conducting,
inviscid fluid described by MHD equations. Initially unmagnetized well-conductive turbulent fluid
generates its own magnetic field which becomes to be dynamically important on almost all relevant
scales. The presence of the large scale field, however, is qualitatively different from the presence
of large-scale flows in hydrodynamics which can be excluded by the choice of reference frame.
The inertial range of MHD turbulence, therefore, has to be dominated by the large-scale
mean magnetic field, which is known as a strong field limit. Initial investigations of
the strong field limit \citep{iroshnikov, kraichnan} prematurely concluded that inertial-range
MHD turbulence has to be weak turbulence, which happened not to be the case. The success
of analytic weak turbulence theories \citep{ng1997,galtier2000} demonstrated that MHD turbulence
has a tendency to become stronger and not weaker during the cascade. Similar arguments
lead \citet{GS95} (thereafter GS95) to conclude that the inertial range of MHD turbulence has to be the so-called
strong critically-balanced anisotropic cascade, which was tentatively confirmed in many earlier
simulations of MHD turbulence, e.g., \citet{cho2000,maron2001}. 

The properties of strong field anisotropic cascade can be rigorously argued to be governed
by the Alfvenic part of MHD perturbations, hence this regime has been dubbed Alfvenic turbulence.
The equations for Alfvenic component called reduced MHD \citep{kadomtsev1974,strauss1976}
has been known in plasma physics for a long time and can be justified based on plasma drift
approximation alone, without resorting to collisions, see, e.g. \cite{schekochihin2009}.
The full compressible MHD also have fast mode cascade \citep{CLV03} which we will not be considered here.
Reduced MHD have an inherent symmetry, similar to hydrodynamic symmetry, which allows
to argue that the power-law scaling of turbulent spectra is indeed possible in the strong mean field and
strong anisotropy limit, e.g., in the inertial range, see, e.g., \cite{B12b}.

While previous numerical work confirmed scale-dependent anisotropy of the strong MHD turbulence,
the precise value of the spectral slope was a matter of debate. As earlier simulations \citep{maron2001,muller2005}
hinted at the $-3/2$ slope, shallower than $-5/3$ predicted by the standard GS95 theory, some
adjustments has been proposed to accommodate this difference \citep{galtier2005,boldyrev2005,gogoberidze2007}.
A model with the so called ``dynamic alignment'' \citep{boldyrev2005, boldyrev2006} has been especially
popular. Our earlier simulations hinted at wider locality of MHD turbulence \citep{BL09b} and also
indicated that the scaling of ``alignment'' tend to flatten out and the spectrum restores its $-5/3$ scaling
at sufficiently high Reynolds numbers (Re) \citep{B11,B12b}. Another challenge was a different scaling of kinetic
and magnetic energies in large-mean field simulation. The magnetic energy is slightly higher
than kinetic energy in the cascade even in the limit of very strong mean magnetic field.
This quantity, residual energy, $RE=E_B-E_v$, goes to zero in the dissipative range, both in simulations
and solar wind data. \cite{muller2005} suggested that it could have a $-2$ scaling, but this is
problematic in the large Re limit, as this would imply unrealistically high difference of magnetic
and kinetic energies at the outer scale, as well as strong non-locality through the whole inertial range.

To carefully investigate these issues we: a) performed simulations of reduced MHD turbulence with resolution up to $4096^3$;
b) while using well-resolved, numerically precise data, also used rigorous quantitative argument known as resolution study
to provide a ``yes/no`` type of test of any hypothesis of universal scaling with a particular power law.

\begin{figure}
\begin{center}
\includegraphics[width=1.0\columnwidth]{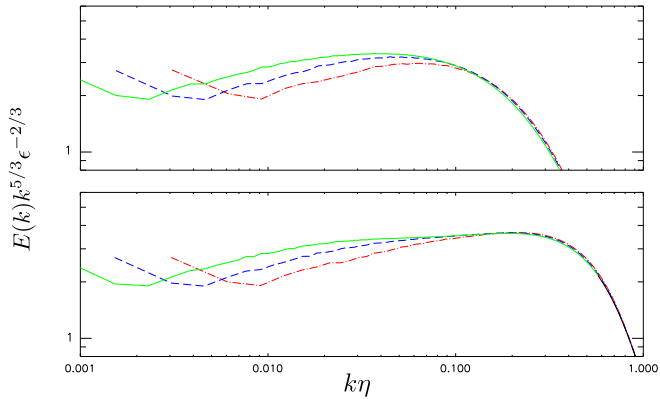}
\end{center}
\caption{Checking -5/3 hypothesis with the scaling study.
Solid, dashed and dash-dotted are spectra from $4096^3$,  $2048^3$ and $1024^3$ simulation correspondingly. The upper plot shows normal diffusion M1-3 simulations
and the lower plot shows hyperdiffusive M1-3H simulations. The convergence is reasonable around the dissipation scale. The scaling that achieves the best convergence is $\approx -1.70$.
The Kolmogorov constant is around 3.5, which is compatible with out previous measurement \citep{B11}.}
\label{energy17}
\end{figure}

\section{Scaling study}
\cite{kolm41} suggested that if strong turbulence is universal and its scaling is only determined by the dissipation rate
and viscosity, the dissipative range would have a certain spacial-, velocity- and time-scales, known as Kolmogorov scales.
This has been tested with a number of experimental and/or numerical data being expressed in units of these scales and presented on the same plot,
see, e.g., \cite{sreenivasan1995, gotoh2002}. This method shows remarkable collapse of all data on the same curve, validating
Kolmogorov's conjecture. Technically, the scaling study investigates the scaling of Kolmogorov dissipation scale
and velocity scale with Reynolds number. For example, in presence of normal viscosity the Kolmogorov velocity should scale
as $Re^{-1/4}$ for the $-5/3$ power law, while for $-3/2$ model it will scale as $Re^{-3/8}$.
The scaling study method becomes especially powerful in numerics, where
all the data are available at all times for averaging.

On a scale $l$ the number of independent realizations in a datacube goes as $l^{-3}$, while the number of correlation timescales for
strong turbulence goes as $l^{-2/3}$, bringing the statistical error down due to averaging by a factor of $l^{-11/6}$, which is about
$10^{-4}$ in highest resolution simulations. The dissipation scale, therefore, is not only the most separated from
the driving scale and the least affected by driving, but also has the smallest statistical error. In combination with very low
numerical error of pseudospectral method (see the subsequent section), the dissipation (Kolmogorov) amplitude of spectra
is one the most robust and best measured quantity in numerics. In particular \cite{kaneda2003}, using simulation group up to $4096^3$
resolution has been able to estimate the power slope of hydrodynamic turbulence within very small error and differentiate
between $-5/3\approx -1.667$ slope and intermittency-corrected $-1.7$. In our work we aim to differentiate between $-3/2$
and $-5/3$ slope, which are different by $\approx 0.167$, much higher than the precision of the method, $\approx 0.02$. 
Other methods, using subjective definition of the inertial range, the ones based on the perceived flatness of the spectra
could easily fail in such task, e.g. due to the transitional scalings that look flat.

The Kolmogorov scale could be expressed as

\begin{equation}
 \eta=(\nu_n^3/\epsilon)^{1/3(n+i+1)},
\end{equation}

where $n$ is the order of dissipation, $i$ is the spectral index, e.g., $-5/3$, $\nu_n$ is viscosity or magnetic diffusivity and $\epsilon$ is the energy
dissipation rate. Checking the hypothesis that the Kolmogorov scale
and the Kolmogorov velocity scales properly with the Reynolds number require plotting the spectrum in Kolmogorov units, i.e.
making the $x$ and $y$ axis dimensionless. The $x$ axis is expressed in $k \eta$, where $\eta$ is not necessarily the classic Kolmogorov scale,
corresponding to $-5/3$ slope, but defined by the above formula, i.e. different for each spectral slope. The $y$ axis
is usually expressed in units of $E(k)k^{5/3+\alpha}L^{\alpha}\epsilon^{-2/3}$, where $\alpha$ is the correction to the $-5/3$ slope
and $L$ is an outer scale, which is normally kept constant in a scaling study. This is, in fact, the only dimensionless expression for
the spectrum that does not contain $\eta$ explicitly. If one wants to multiply the above expression by some power of $(L/\eta)$
or the Reynolds number, this would introduce explicit $\eta$ dependence and would violate the so-called zeroth law of turbulence
which claims that large-scale properties are largely independent on viscosity.

\begin{table}
\begin{center}
\caption{Three-dimensional RMHD simulations}
  \begin{tabular*}{1.00\columnwidth}{@{\extracolsep{\fill}}c c c c c c}
    \hline\hline
Run  & $N^3$ & Dissipation & $\langle\epsilon\rangle$ &  $k_{\rm max}\eta$ & $k_{\rm max}\eta_B$ \\
   \hline
M1 & $1024^3$ & $-1.75\cdot10^{-4}k^2$   & 0.06 &  1.05 & 2.00 \\
M2 & $2048^3$ & $-7\cdot10^{-5}k^2$  & 0.06  &  1.06 & 2.17\\
M3 & $4096^3$ & $-2.78\cdot10^{-5}k^2$ & 0.06 &  1.06 & 2.34\\
\hline
M1H & $1024^3$ & $-1.6\cdot10^{-9}k^4$   & 0.06 &  1.04 & 1.37\\
M2H & $2048^3$ & $-1.6\cdot10^{-10}k^4$  & 0.06  &  1.04 & 1.42\\
M3H & $4096^3$ & $-1.6\cdot10^{-11}k^4$ & 0.06 &  1.04 & 1.47\\
   \hline
\end{tabular*}
  \label{experiments}
\end{center}
\end{table}

\begin{figure}[b]
\begin{center}
\includegraphics[width=1.0\columnwidth]{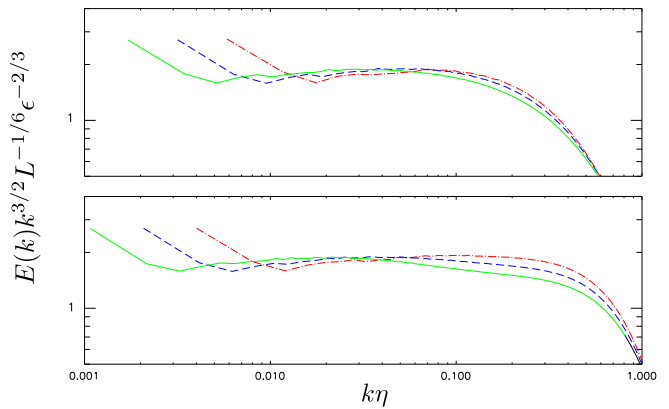}
\end{center}
\caption{Checking -3/2 hypothesis with the resolution study. The convergence is poor. Convergence is required starting with dissipation scales, as the opposite
would mean that either of the Kolmogorov scales depends differently on Re, as was expected. Note that there's no convergence for the lower wavenumbers
either, contrary to what was claimed in \citet{perez2012}. This was due to the scaling near driving scale being shallower than $-3/2$, around $-1.4$. Such
scaling near the driving scale is always affected by driving and the fine-tuning of driving is required to achieve $-3/2$ slope.}
\label{energy15}
\end{figure}

\section{Numerics}
We performed two series of reduced MHD driven simulations with a strong mean field $B_0$ in code units, RMS fields $v_{\rm rms}\approx B_{\rm rms} \approx 1$,
perpendicular box size of $2\pi$ and parallel box size of $2\pi B_0$. The driving was correspondingly anisotropic with anisotropy $B_0$, so that turbulence starts
being strong from the outer scale. Our previous simulations \citep{B12b} showed rapid decrease of parallel correlation length right after the driving scale, which indicates
efficiency of nonlinear interaction and the strong turbulence regime. The correlation timescale for $v$ and $B$ was around $\tau\approx 0.97$, so the box contained roughly
$6.5$ parallel correlation lengths in parallel direction and about $3-5$ in perpendicular direction. Each simulation was started from long-evolved low resolution simulation,
and was subsequently evolved for $\Delta t=13.5$ in high resolution. Overall, our setup is very similar to our previous simulations \cite{B11,B12b} with the exception for driving
that was limited to lower $k<1.42$ wavenumbers in this simulations. The reader is welcome to study our method in more detail using the above references.
We used the last $7$ dynamical times for averaging. In our previous simulations we found that averaging
over $\sim 7$ correlation timescales gives reasonably good statistic on outer scale and excellent statistics on smaller scales (see the above estimates).
The simulation parameters are listed in the Table 1. Numerically, we used $k_{\rm max} \eta>1$ resolution criterion, with $\eta$ being classic Kolmogorov scale,
 that was shown to be sufficient in normal viscous
simulations, e.g. \cite{gotoh2002} and was a better resolution that the one used in \cite{perez2012}.

For hyperdiffusive series we used the same criterion, additionally
we checked numerical precision of the spectra by performing resolution study on lower resolutions. In particular we saw spectral error lower than $8\times 10^{-3}$
up to $k \eta =0.5$ when increasing resolution from $576^3$ to $960^3$ and the spectral error lower than $3\times 10^{-3}$ when we increased parallel
resolution in a $1152^3$ simulation by a factor of two. We didn't use any data above $k \eta =0.5$ for fitting as the spectrum sharply decline after this point and
contains negligible energy. 
We conclude that for our purposes using $k_{\rm max} \eta=1$ is sufficient and using cubic resolution, i.e. parallel resolution equal to perpendicular resolution
is also sufficient or even somewhat excessive. Note that increasing resolution while keeping $k_{\rm max} \eta>1$ with $\eta$ corresponding to $-5/3$
slope is a conservative choice for all types of turbulence with slope shallower than $-5/3$, including the $-3/2$ model. Table~1 also lists $k_{\rm max}\eta_B$
with $\eta_B$ corresponding to $-3/2$ model.

\begin{figure}
\begin{center}
\includegraphics[width=1.0\columnwidth]{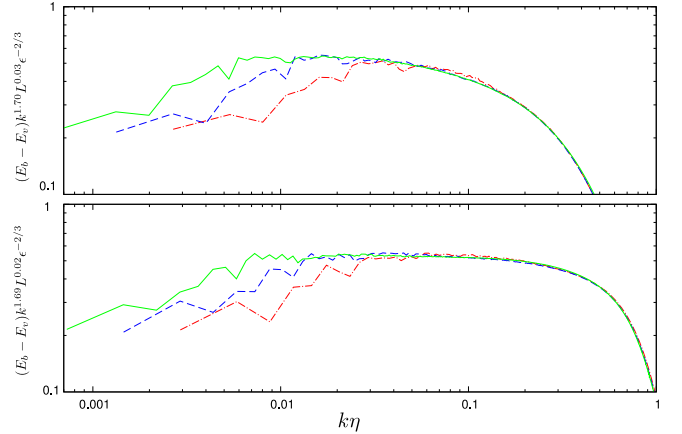}
\end{center}
\caption{Residual energy convergence. Best convergence is $k^{-1.70}$ scaling for M1-3 and $k^{-1.69}$ scaling for M1-3H. }
\label{res}
\end{figure}

\section{Discussion}
Our results suggest that the residual energy scales as the total energy and is simply a constant fraction of the total energy. Our best estimate for this fraction
is $0.15\pm0.03$. The discussion of the fraction of the residual energy and its scale-dependence dates back a couple of decades and has recently been connected
to other dimensionless measures called alignment measures in simulations \citep{BL09b} and in solar wind measurements \citep{wicks2013}. Technically,
the puzzle of scale-dependence of the fraction of residual energy is as important as the question of scale-dependency of a particular measure introduced by
\citet{mason2006}, which has been claimed many times by \citet{mason2008,perez2012} to be the exact measure that reduces nonlinear interaction
and is responsible for the modification of the spectral slope. In our previous work we found that although ''dynamic alignment`` slope indeed somewhat correlates
with the correction to the spectral slope, this relation is not exact. We are also not aware of any convincing theoretical argument why ''dynamic alignment`` should be
preferred to other alignment measures, see \citet{BL09b} for a discussion. The recent results that in higher Re simulations the alignment measures become
constant and the slope approaches Kolmogorov slope significantly simplify the picture and makes the discussion of the alignment influence to
the nonlinear cascade largely irrelevant. Explaining previously suggested $-2$ scaling \citep{muller2005} for the residual energy is especially challenging
in theory as, combined with the fact that the fraction of residual energy is fairly small near the dissipation scale both in numerics and in the solar wind,
the $-2$ scaling would suggest arbitrarily large fraction of the residual energy with sufficiently high Re. This would be seriously at odds with our understanding
of MHD turbulence in the presence of strong mean field. Our work confirming that the residual energy is likely to be just a fraction of the total energy
resolve this conceptual difficulty and make theories suggesting different scalings for magnetic and kinetic energies unnecessary. While the practically
measured spectra, such as solar wind spectra usually feature different kinetic and magnetic scalings these does not necessarily directly suggest for
the changes in theory for the asymptotic behavior at very high Re. 

Using scaling study is relatively new in MHD and the first tentative confirmation of $-5/3$ slope was published in \cite{B11}. This paper was heavily criticized
in \citet{perez2012} for being numerically unresolved in parallel direction. This criticism was misguided, however, as high numerical accuracy is not required for the
scaling study argument, involving confirmation of a particular scaling as long as $\eta$ for this particular model scales precisely with the grid scale. This is because
the numerical error on grid scale depends only on $k_{\rm max}\eta$ and the possible distortion of the spectra will be exactly the same as a function of $k\eta$
for each simulation. Therefore as long as the hypothesis is correct and the scaling correspond to this hypothesis the numerical spectra will still collapse onto the
same curve. Logically, however, this argument does not work if one wants to reject a hypothesis of a different scaling because if the scaling is different, 
the $k_{\rm max}\eta$ will be different for each simulation and the numerical error that correspond to it, or to the unresolved parallel direction, will be
different. This being said, in \cite{B11}  $k_{\rm max}\eta$ was close to unity for either model, owing to the high order of the dissipation term. So, in practicality
\cite{B11} also rejected the $-3/2$ scaling. The subsequent work \citet{B12b} actually contained fully resolved simulations, such as R4-5, which conclusively
rejected the $-3/2$ scaling. In this paper we have opted to perform fully resolved, numerically accurate simulations in order to avoid delving into such
complicated matters. However, in our opinion \cite{B11, B12b} conclusively and rigorously supported $-5/3$ scaling and rejected $-3/2$ scaling
for high Re simulations. The other part of the criticism of \citet{perez2012}, that \citet{B11} measured spectral slopes that were distorted
due to hyperdiffusion, was the result of a misunderstanding of the scaling study argument. The scaling study does not measure any particular
slope at any point of the numerical spectrum, instead, as we explained in \citet{B12b} and this Letter, it measures how $v_{\rm eta}$ and $\eta$
scale with Re. Such scalings are expected to be universal for high Re and are insensitive to the type of dissipation that was used.
A simple way to confirm this is to formulate the scaling convergence in terms of a different type of spectrum, e.g. 1D spectrum. It is easy to show
that if convergence is present for 3D spectrum, it will also be present for 1D spectrum as well, despite 1D spectrum have very different spectral
distortions due to the bottleneck effect. This has been well known since long time ago in hydrodynamics as both 1D and 3D spectra has been
used for scaling studies, e.g., \cite{gotoh2002}. One comment is in order, however. The visible $\approx -1.4$ scaling in the beginning of the spectrum
is real in a sense that if one makes a scaling study with lower Re, e.g. with all simulations having $Re<2000$, it will confirm the $-1.4\div -1.5$ scaling.
This should not be surprising as the universal scaling is only expected for very high Re.

The simulations presented in this paper and also in \citep{B11, B12b} are using the same equations, similar box size prescription and large-scale
driving prescription to the once used in \citet{perez2012}. We are not aware of any significant differences in terms of raw spectra
\footnote{Despite their simulations were slightly under-resolved
by the $k_{\rm max}\eta$ criterion compared to ours.} presented by their group
and us, except for the data anomaly for the highest resolution spectrum on Fig.~8 in \citet{perez2012}, see \citet{B13comm}.
What is the source of the radically different claims about the spectral slope between these works? Firstly, theirs are
lower-resolution data (up to $2048^3$, vs ours $4096^3$). Secondly, the claim of convergence on Fig.~8 in \citet{perez2012} is in visible
odds with the figure itself, i.e. the convergence is indeed absent for the $-3/2$ slope. There was also a number of logically incorrect and misleading
statements regarding the length of the inertial range, purportedly confirming the Boldyrev scaling, but instead being a logical loop argument \citep{B13comm}.

To summarize, the highest resolution MHD simulations to-date, with Re up to $36000$ exhibit asymptotic spectral scaling of around $-1.7$, slightly steeper than Kolmogorov.
The residual energy and also kinetic and magnetic energies separately exhibit the same scaling.

\section{Acknowledgements}
This work was supported by the DOE INCITE program. The computations for M1-3 and M1-3H were performed on the ALCF Intrepid cluster, using approximately 20 million CPU hours.
The numerical results of our imbalanced simulations supported by the same project will be presented in a subsequent publication.
We thank our ALCF Catalyst Tim Williams for the technical help. Our special thanks go to Dmitry Pekurovsky, who wrote a customized version of his P3DFFT
library \citep{Pekur2012} which made $4096^3$ MHD simulations possible. The full time evolution for 1.2 correlation timescales for the $2048^3$ simulations and several datacubes
for the $4096^3$ simulations will be publicly released in the Johns Hopkins University turbulence database http://turbulence.pha.jhu.edu/ at a future date.

\

\def\apj{{\rm ApJ}}           
\def\apjl{{\rm ApJ }}          
\def\apjs{{\rm ApJ }}          
\def\grl{{\rm GRL }}
\def\aap{{\rm A\&A } }
\def\mnras{{\rm MNRAS } }
\def\physrep{{\rm Phys. Rep. } }               
\def\prl{{\rm Phys. Rev. Lett.}} 
\def\pre{{\rm Phys. Rev. E}} 
\bibliography{all}

\end{document}